\documentclass{openjournal}

\usepackage{xcolor}
\usepackage{textgreek}
\usepackage[utf8]{inputenc}
\usepackage[english]{babel}
\usepackage{amsmath}
\usepackage{hyperref}
\hypersetup{
    unicode, 
    colorlinks=true,
    linkcolor=linkcolor,
    citecolor=linkcolor,
    filecolor=linkcolor,
    urlcolor=linkcolor,
}
\usepackage{color,colortbl}
\definecolor{linkcolor}{rgb}{0.0,0.3,0.5}
\usepackage{tensind}
\tensordelimiter{?}
\DeclareGraphicsExtensions{.bmp,.png,.jpg,.pdf}
\usepackage{verbatim}
\usepackage[normalem]{ulem}
\usepackage{orcidlink}
\usepackage{soul}

\newcommand{\given}{\, | \,}
\newcommand{\btheta}{\boldsymbol{\theta}}
\newcommand{\by}{\boldsymbol{y}}

\newcommand{\bmu}{\boldsymbol{\mu}}
\newcommand{\bC}{\boldsymbol{C}}

\newcommand{\revision}{}

\makeatletter
\long\def\@makecaption#1#2{%
  \vskip\abovecaptionskip
  \small
  \noindent #1: #2\par
  \vskip\belowcaptionskip
}
\renewcommand{\fnum@figure}{FIGURE~\thefigure}
\renewcommand{\fnum@table}{TABLE~\thetable}
\makeatother

\graphicspath{ {./figs/} }

\begin{document}

\setlength{\parskip}{0.3\baselineskip}

\title{The Information Content of Quasar Variability Light Curves: \\
How Well Can we Infer Stochastic Model Parameters?\vspace{-2.5em}}

\author{Brendon J. Brewer\footnote{\href{mailto:bj.brewer@auckland.ac.nz}{bj.brewer@auckland.ac.nz}}\ \orcidlink{0000-0001-9902-7112}}
\affiliation{Department of Statistics, The University of Auckland, Private Bag 92019, Auckland 1142, New Zealand}

\author{Geraint F. Lewis \orcidlink{0000-0003-3081-9319}}
\affiliation{Sydney Institute for Astronomy, School of Physics, A28, The University of Sydney, NSW 2006, Australia}

\author{Xiang Yu (Ryan)}
\affiliation{Department of Statistics, The University of Auckland, Private Bag 92019, Auckland 1142, New Zealand}

\author{Yuan Li (Cher)}
\affiliation{Department of Statistics, The University of Auckland, Private Bag 92019, Auckland 1142, New Zealand}


\begin{abstract}
\vspace{1em}
\begin{center}
ABSTRACT \\ 
\end{center}
Quasar variability, driven by multi-scale physical processing within a relativistic accretion disk, is commonly modelled with stochastic time series
models. The simplest of these is the
Damped Random Walk (DRW),
also known as the Ornstein-Uhlenbeck (OU) process.
Here, we demonstrate that, when fitting such a model to quasar light
curve data, the mean of the light curve, $\mu$, should not be fixed (which is the typical approach), as this leads to overconfident inferences
about the variability timescale $\tau$, with substantially underestimated
uncertainties.
However, the short term volatility parameter
$\eta$ is typically very well constrained from short light curves.
Through simulations, we compute information theoretic quantities
such as the conditional entropy and the mutual information,
confirming that light curves provide much more information about
$\eta$ than about $\tau$.
As a result, we recommend that future
quasar variability studies focus on $\eta$ rather than $\tau$.
To demonstrate this approach, we fit a
hierarchical Bayesian regression
model for $\eta$ as a function of bolometric luminosity and rest
wavelength to a dataset of 570 light curves measured over decades.
We perform the fit using a likelihood function that uses
the light curves
directly, rather than using intermediate $\eta$ values from individual
light curve fits. We find that volatility decreases as a function
of both
bolometric luminosity and rest wavelength.
The volatility also decreases more steeply with redshift than
time dilation alone would suggest, pointing to an increase in
intrinsic volatility as quasars evolve over cosmic time.
\end{abstract}





\section{Introduction}
\label{sec:intro}
Quasars are among the most luminous and dynamic objects in the Universe, their prodigious energy output powered by the accretion of matter onto supermassive black holes residing in the centres of galaxies \citep[see][]{Padovani_2017}. They exhibit variability across essentially all observed wavelengths and on timescales ranging from hours to decades. This variability is widely understood to arise from the intricate and highly non-linear physics operating within relativistic accretion discs, where turbulent magnetohydrodynamic flows, radiation transport, thermal instabilities, and strong gravitational effects combine to produce complex and evolving emission signatures \citep[e.g.][]{Czerny_2023}. The resulting variability provides a unique observational window into physical processes occurring on scales otherwise inaccessible to direct imaging.

Given the physical complexities, {\it ab initio} modelling of quasar light curves presents a significant challenge \citep[e.g.][]{Fagin_2025}. Hence, quasar variability is 
often represented as a statistical process described by a few parameters.
It has been established that a Damp Random Walk (DRW) approach, in particular the CAR(1) process (continuous autoregressive process of order 1),
represents quasar variability well \citep{kelly2009variations}. Whilst there is some evidence that
the quasar variability might depart from this model
\citep[e.g.][]{zu2013quasar}, with more sophisticated models being available \citep{kelly2014flexible}, it remains a useful approximation with a relatively simple set of parameters including a characteristic timescale $\tau$.
However, to accurately determine the CAR(1) parameters, sufficiently long temporal observations of quasar light curves are essential, with durations exceeding the typical time scale by a factor of at least ten \citep{kozlowski2017limitations}. However, the statistical characteristics of variability feed into other quasar studies, such as reverberation mapping and using quasars as cosmological probes.

Extensive samples of quasar light curves are now becoming available, with \citet{stone}, hereafter S22,
presenting multiband variability and estimates of $\tau$ (with uncertainties)
for a sample of almost 200 quasars over a two decade period.
Assuming that the timescale parameter $\tau$ can be used as a clock to time quasar variability, \citet{lewis2023detection}
and \citet{brewer2025revisiting} used the \citet{stone} estimates
to confirm the expected presence of cosmological time dilation in the variability timescales. However, the robustness of this cosmological measurement is dependent upon robustness of the measurement of the timescale $\tau$, and hence it is essential to
investigate the reliability of the inference processes involved in determining
$\tau$ from the light curves.
This will form the focus of this paper.

The layout of this paper is as follows: Section~\ref{sec:fitting} covers the background required for fitting CAR(1) models including details about the parameters, the prior
distributions, the likelihood function, and the computational approach.
In Section~\ref{sec:single}, we demonstrate the issues involved in fitting
CAR(1) models to light curve data, based on a simulated dataset
as an example.
In Section~\ref{sec:information}, we use information theory to quantify the
amount of information light curves typically contain about the stochastic model
parameters. Section~\ref{sec:scoring} extends this analysis to quantify the
cost of certain modelling choices in terms of the reliability of the inferences
about $\tau$. The results of these sections lead us to focus on a short term
volatility parameter $\eta$ as it is much better constrained than $\tau$.
In Section~\ref{sec:hierarchical} we fit a hierarchical regression model to
the S22 sample of quasars to determine how quasar volatility $\eta$
depends on redshift, luminosity, and the rest wavelength of the observations.
Finally, we conclude in Section~\ref{sec:conclusion}.

\section{Bayesian CAR(1) Fitting}\label{sec:fitting}
Typical light curve data for a particular quasar in a particular waveband
consists of a series of timestamps
$\boldsymbol{t} = \{t_1, ..., t_N\}$, a corresponding sequence of brightness measurements
$\by = \{y_1, ..., y_N\}$ (which we will take to be in magnitudes),
and error bars $\boldsymbol{s} = \{s_1, ..., s_N\}$ which indicate the
uncertainty on the brightness measurements.
Our notation for the magnitude measurements uses a subscript (e.g., $y_i$) for
the observed values including measurement error.
When we discuss the underlying error-free
continuous curve, we still use the symbol $y$ but with parentheses; i.e.,
$y(t)$.

\subsection{Bayesian Inference}
To fit a stochastic variability model to such data, we must
identify the model parameters $\btheta$,
specify their prior distribution $p(\btheta)$,
and also specify the sampling distribution\footnote{Sampling distribution
is the traditional term, but the sampling distribution is
best understood as a representing
prior information about the relationship between the data
and the parameters \citep{caticha2006updating, caticha2021entropy}.} for the data
$p(\by \given \btheta)$, describing what
data we would expect to observe as a function of the parameters.
Bayes's theorem then gives the posterior distribution for the
model parameters
\begin{align}
p(\btheta \given \by)
    &= \frac{p(\btheta)p(\by \given \btheta)}
            {p(\by)},
\end{align}
where $p(\by \given \btheta)$ is the likelihood function (obtained
by substituting the observed data into the expression for the sampling
distribution) and the denominator is the marginal likelihood or evidence
value, given by
\begin{align}
p(\by) &= \int p(\btheta)p(\by \given \btheta) \, d\btheta.\label{eqn:marginal_likelihood}
\end{align}
The posterior distribution is often characterised by generating samples
from it using Markov Chain Monte Carlo (MCMC) techniques
\citep{sharma2017markov}. If required, marginal likelihoods and
posterior samples can also be computed simultaneously using
Nested Sampling \citep{skilling}.
In this paper, we use Diffusive Nested Sampling \citep{dns},
as implemented in the DNest4 package \citep{dnest4},
to sample from posterior distributions and compute marginal likelihoods.

\subsection{CAR(1) Likelihood Function}
The likelihood function for the CAR(1) model can be obtained in
two equivalent ways --- using the state space representation
\citep{kelly2009variations} or using Gaussian processes
\citep{williams2006gaussian}. We will use the latter framework
as it is more familiar to most readers and to the authors.
For a review of Gaussian process applications in astronomy, see
\citet{aigrain2023gaussian}.

In the CAR(1) model, the joint distribution for the signal
at a collection of times $t_1, ..., t_N$, conditional on the
CAR(1) parameters, is a multivariate normal distribution, given
by
\begin{align}
p(\by \given \btheta)
    &= \frac{1}{\sqrt{(2\pi)^N \det \bC}}
        \exp\left(-\frac{1}{2}(\by - \bmu)^T\bC^{-1}(\by - \bmu)\right).\label{eqn:gp}
\end{align}
The CAR(1) model implies a certain form for the covariance
matrix $\boldsymbol{C}$, which depends on the CAR(1) parameters.
Specifically,
the covariance between the signal at two different times,
$y(t_1)$ and $y(t_2)$, is given by
\begin{align}
C(t_1, t_2) &= \sigma^2 \exp\left(-\frac{|t_2 - t_1|}{\tau}\right),\label{eqn:covariance_function}
\end{align}
where $\tau$ is a timescale parameter describing how long it takes
for $y(t)$ to be somewhat less correlated with its previous values,
and $\sigma$ is a parameter
describing the typical degree of deviation of $y(t)$ from its mean
value $\mu$ (which we assume to be a constant with time ---
making $\boldsymbol{\mu}$ in Equation~\ref{eqn:gp} a vector of length
$N$, all of whose elements are the same value).
Equation~\ref{eqn:covariance_function} gives the covariance
for the underlying measurement-error-free signal $y(t)$ at
the timestamps. However, we also need to take into account measurement
error, in order for Equation~\ref{eqn:gp} to be valid for the
observational data itself.
To achieve this, we add extra values to the diagonal of the
covariance matrix, based on the reported error bars
$\{s_i\}$ and
an additional white noise parameter, commonly called `jitter'
(which can also be interpreted
as expanding the error bars by an amount
that is added in quadrature to the reported error bars).
The covariance matrix elements are therefore
\begin{align}
C_{ij} &= \sigma^2 \exp\left(-\frac{|t_j - t_i|}{\tau}\right)
                + \left[s_i^2 + (\textnormal{jitter})^2\right]
                    \delta_{ij},
\end{align}
where $\delta_{ij}$ is the Kronecker delta.
In practice, it is computationally expensive to evaluate
the likelihood function of Equation~\ref{eqn:gp}.
Naive calculation is very expensive, and ordinarily it
can be significantly sped up using the Cholesky
factorisation of the covariance matrix $\boldsymbol{C}$.
However, this approach still scales as $O(N^3)$.
For certain types of covariance matrix such as the CAR(1)
covariance function,
specialised acceleration techniques are possible
due to the sparse and/or banded structure of the inverse
covariance matrix, reducing the computational cost to
$O(N)$. To benefit from these speed-ups
we used the {\tt celerite2} Python package \citep{celerite2}
to implement the log likelihood function throughout this study.

The parameter $\sigma$ has a simple interpretation as the stationary
(marginal)
standard deviation of $y(t)$ at all times, and thus describes how
different $y(t)$ is from its typical value $\mu$, over long
timescales. However, when inferring the parameters
$\btheta = \{\mu, \sigma, \tau, {\rm jitter}\}$ from
light curves using this model, there is often a very strong
correlation between $\sigma$
and $\tau$ in the posterior distribution. Datasets can often 
be explained by either low $\sigma$ and low $\tau$, or high
$\sigma$ and high $\tau$.
To mitigate any MCMC sampling inefficiencies caused by this correlation,
it is usually better to work in terms of a short term volatility parameter
given by
\begin{align}
\eta &= \sigma\sqrt{\frac{2}{\tau}}.\label{eqn:eta}
\end{align}
This $\eta$ parameter is the one that naturally appears in the
stochastic differential equation formulation of this
model \citep{kelly2009variations}.
For completeness, the stochastic differential equation
for this process is given by
\begin{align}
dy(t) &= -\frac{1}{\tau}(y(t)- \mu) \, dt + \eta \, dW(t), 
\end{align}
where $W(t)$ is a standard Wiener process and $dW(t)$ is a
normally distributed white noise increment.
When we work with days as the time units, $\eta$ has a simple
interpretation, as long as $\tau \gg 1$ day --- it is the
typical size of day-to-day fluctuations in $y(t)$.
In summary, the parameters of the model for a single light
curve are:
\begin{itemize}
\item $\mu$, the typical magnitude around which the light
curve fluctuates;
\item $\sigma$, the degree of variation around $\mu$;
\item $\tau$, the timescale over which the signal $y(t)$
becomes decorrelated from its past values;
\item $\eta$, the short term volatility; and
\item ${\rm jitter}$, the amount of extra independent noise
that is present, above and beyond what is reported in the
error bars $\{s_i\}$.
\end{itemize}
Of the middle three parameters, only two are actually
genuine parameters and the third may be derived from the other
two. The choice of which parameters to use as coordinates is
discussed in detail in Section~\ref{sec:choice}.

Another derived parameter that may be of interest is a measure
of annual variability. We can compute the conditional distribution
of the signal $y(t+365)$ given the current signal $y(t)$ and the
parameters~\footnote{Note that all temporal properties are given in units of days throughout this paper unless explicitly noted.}. This is a normal distribution
whose expected value represents probable drift towards
$\mu$:
\begin{align}
y(t+365) \given y(t) &\sim
    \textnormal{Normal}\left(\mu + (y(t)-\mu)\exp(-365/\tau),
                            \sigma^2(1 - \exp(-(2\times 365)/\tau))\right).
\end{align}
The standard deviation is simply the square root of the variance,
which, unlike the mean, does not depend on the current value of the
signal $y(t)$. It can be used straightforwardly as a measure of
annual variation:
\begin{align}
\textnormal{sd}\left(y(t+365) \given y(t)\right) &= 
    \sigma\sqrt{(1 - \exp(-(2\times 365)/\tau))}.
\end{align}

\subsection{Choice of Coordinates and Prior Distributions}\label{sec:choice}
In this paper we will use two different prior distributions
for the CAR(1) parameters: a somewhat informative, physically motivated prior,
and an uninformative, convenient flat prior (flat in the logarithm for
some parameters). The issues we explore are present
in both cases, but tend to be more dramatic in the case of the flat prior.
However, before specifying the priors, we must take care to choose an
appropriate coordinate system on the parameter space
for implementing the inference. Using
$(\mu, \sigma, \tau, {\rm jitter})$ as the parameters, as discussed
previously, is intuitively clean, but often
results in very strongly dependent posterior distributions (especially
between $\sigma$ and $\tau$), which can cause MCMC mixing problems.
To avoid this, $(\mu, \eta, \tau, {\rm jitter})$ is often preferred,
where the stationary standard deviation $\sigma$ has been replaced
by the short term volatility $\eta$.
The stationary standard deviation
$\sigma$ can then be calculated as a derived parameter using
Equation~\ref{eqn:eta}, if it is of interest.

However, we have elected to use an unconventional third option, parameterising
via $(\mu, \sigma, \eta, {\rm jitter})$. This retains the computational
advantages of $(\mu, \eta, \tau, {\rm jitter})$ (i.e., that the joint posterior
tends to be fairly independent), while also allowing us to
easily specify a prior for $\sigma$ since it is one of the
coordinates (in our view, it is easier to think about plausible
values of $\sigma$, which describes how much an individual quasar
varies, than it is to think about plausible values of $\tau$).
In this parameterisation,
$\tau$ becomes a derived parameter, given by
\begin{align}
    \tau &= 2\left(\frac{\sigma}{\eta}\right)^2,
    \label{eqn:tau}
\end{align}
which is simply a rearrangement of Equation~\ref{eqn:eta}.
Note that the issues presented in Sections~\ref{sec:single},~\ref{sec:information},
and~\ref{sec:scoring} are not
caused by this choice of coordinates, and
remain present no matter which parameterisation is used.

We now turn our attention to the prior distribution for the
parameters $(\mu, \sigma, \eta, {\rm jitter})$.
Taking the logarithm of both sides of Equation~\ref{eqn:tau} gives\footnote{Throughout this paper, $\log()$ denotes the natural logarithm
and $\log_{10}()$ is the base-10 logarithm.}
\begin{align}
\log_{10}(\sigma) &= \log_{10}(\eta) + \frac{1}{2}\log_{10}(\tau) - \frac{1}{2}\log_{10}(2).
\end{align}
which is a linear transformation
from $\log_{10}(\eta)$ and $\log_{10}(\tau)$ to $\log_{10}(\sigma)$.
Therefore, if we were to assign
a joint normal distribution for the logarithmic quantities
in one of the parameterisations,
we would also have a joint normal distribution in the other ---
a convenient property. Normal distributions thus seem to be a
natural and convenient choice for the informative priors.
However, in some Nested Sampling implementations,
the parameters are ultimately encoded as coordinates with
uniform prior distributions between 0 and 1, and the
inverse of the cumulative distribution function (CDF)
is then used to transform the coordinates to the actual model
parameters with the appropriate prior\footnote{DNest4
does not require this structure in general. However, the Python
model implementation feature of DNest4, which we use in this study,
does require this structure.}. For the normal
distribution this operation can be computationally costly as it needs to
occur millions of times.
Therefore, we chose to use the `logistic'
distribution in place of the normal distribution.
The logistic distribution is a good approximation to the normal distribution
(but with slightly heavier tails),
and has a fast inverse CDF which is simply $\log(u/(1-u))$.
The informative priors we chose are:
\begin{align}
    \mu &\sim \textnormal{Logistic}({\rm mean}=20, {\rm sd}=1.5) \\
    \log_{10}(\sigma) &\sim \textnormal{Logistic}({\rm mean}=-0.5, {\rm sd}=0.5) \\
    \log_{10}(\eta) &\sim \textnormal{Logistic}({\rm mean}=-2, {\rm sd}=0.6) \\
    \log_{10}({\rm jitter}) &\sim \textnormal{Logistic}({\rm mean}=-2, {\rm sd}=0.5).
\end{align}
For ease of thinking, we have written the logistic distributions in terms
of their means and standard deviations. These are very close approximations
to normal distributions with the same means and standard deviations,
but with the computational advantage of a cheaper inverse CDF.
The means and standard deviations of the informative priors were chosen using the following
arguments.
The informative prior for mean magnitude $\mu$ is loosely based on the frequency distribution
of magnitudes
of quasars observed in large surveys \citep[e.g.,][]{10.1093/mnras/staf1398}.
The prior for the long term variability standard deviation parameter $\sigma$ was
set to make it quite likely
(approximate 95\% prior credible interval)
to be between $0.0316$ and $3.16$ magnitudes.
The prior for the short term volatility $\eta$ was selected to make it
very likely to be between
$0.000631$ and $0.158$ magnitudes ${\rm day}^{-1/2}$.
The interpretation of $\eta$ is straightforward (at least when $\tau$
is longer than a day), as it is the typical
variation in magnitude seen from one day to the next. The
power of $-1/2$ in the units arises because these
changes accumulate over time as a random walk, rather than linearly.
Finally, the jitter prior makes it very likely to be between
0.001 and 0.1 magnitudes.

On the other hand,
the flat priors we chose are given by
\begin{align}
    \mu &\sim \textnormal{Uniform}(15, 25) \\
    \log_{10}(\sigma) &\sim \textnormal{Uniform}(-3, 1) \\
    \log_{10}(\eta) &\sim \textnormal{Uniform}(-5, 0) \\
    \log_{10}({\rm jitter}) &\sim \textnormal{Uniform}(-3, -1).
\end{align}
The range for the mean magnitude $\mu$ comfortably covers a plausible range
for quasars, with some leakage into implausible values, which is common
with flat priors. For the long term standard deviation parameter $\sigma$,
we allow it to be between a very small lower limit of $10^{-3}$ magnitudes
and a very large upper limit of $10$ magnitudes, with a log-uniform prior.
For the short term volatility parameter $\eta$, the prior allows for
values down to $10^{-5}$ (very tiny magnitude deviations over $\sim$ days),
up to $1$ (extreme magnitude deviations on a daily basis), with a log-uniform
prior. The jitter is allowed to be between $10^{-3}$ magnitudes and 0.1
magnitudes, again with a log-uniform prior.

\begin{figure}
\begin{center}
\includegraphics[width=0.5\textwidth]{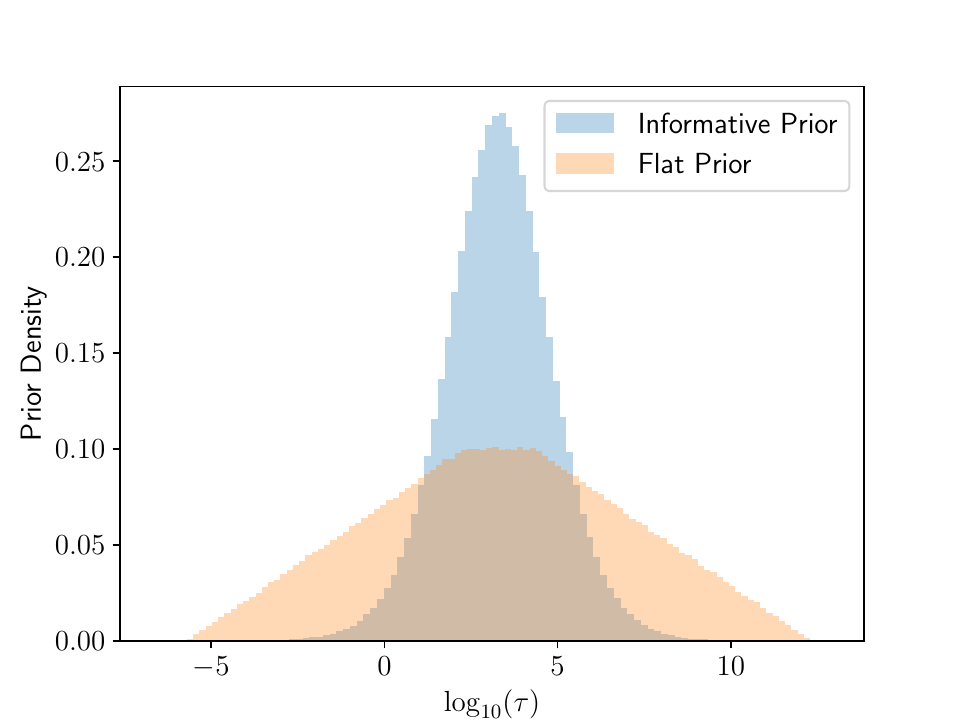}
\end{center}
\caption{The induced prior distributions for $\log_{10}(\tau)$
(with $\tau$ measured in days)
with the informative priors (very close to a normal distribution) and the
flat priors. Both priors cover several orders of magnitude,
with the flat prior case being wider, as expected. For reference,
$10^5$ days is 273 years, and $10^{10}$ days is 27.3 million years.
\label{fig:tau_priors}}
\end{figure}

The induced priors for $\log_{10}(\tau)$, now a derived parameter, in the informative and flat cases are shown in Figure~\ref{fig:tau_priors}.
In the informative case, the prior is very close to a normal
distribution (if our informative priors had been normal, this one
would be too, and our logistic priors approximate this).
Its mean is about 3.3 (corresponding to a prior median of about
2000 days) and its standard deviation is about 1.6 dex.
In the flat prior case, the derived prior for $\log_{10}(\tau)$
is broader and is, in fact, a trapezoidal shape. Its mean is also about 3.3
and its standard deviation is about 3.7 dex.
There is some leakage of the $\log_{10}(\tau)$
prior to implausibly small (sub-day) values, but this often occurs
with conservative prior choices. Very large values above a million
years ($\log_{10}(\tau) \gtrsim 8.6$) are possible with a modest
prior probability.
{\revision The wide priors for $\tau$ that we use throughout this
paper include possible values that are very large compared to
both the typical duration of measured light curves, and
compared to
estimates that are typically reported.
However, if $\tau$ is truly measurable from the light curves,
very large values will be ruled out in the
posterior distribution.}

\vspace{10mm}

\section{The Posterior Distribution from a Single Light Curve}\label{sec:single}
We will now demonstrate the main difficulties with inferring
the CAR(1) parameters $\btheta = \{\mu, \sigma, \eta, {\rm jitter}\}$,
and hence the derived timescale parameter $\tau$, from the data $\by$.
A key issue in this kind of inference is whether the parameter
$\mu$ is fixed at an estimated value (or equivalently, set to zero
after subtracting an estimated value from $\by$),
or remains a truly free parameter.
If it is fixed at an estimated value, this is often the arithmetic mean
\begin{align}
\hat{\mu} &= \frac{1}{N}\sum_{i=1}^N y_i
\end{align}
or a weighted mean using the error bars
\begin{align}
\hat{\mu} &= \frac{\sum_{i=1}^N \left(y_i/s_i^2\right)}
                  {\sum_{i=1}^N \left(1/s_i^2\right)}.\label{eqn:fixed_mu}
\end{align}
Note that fixing $\mu$ to an estimate $\hat{\mu}$ is equivalent to
subtracting an estimate from $\by$ and then assuming $\mu = 0$.

\begin{figure}
\begin{center}
\includegraphics[width=0.7\textwidth]{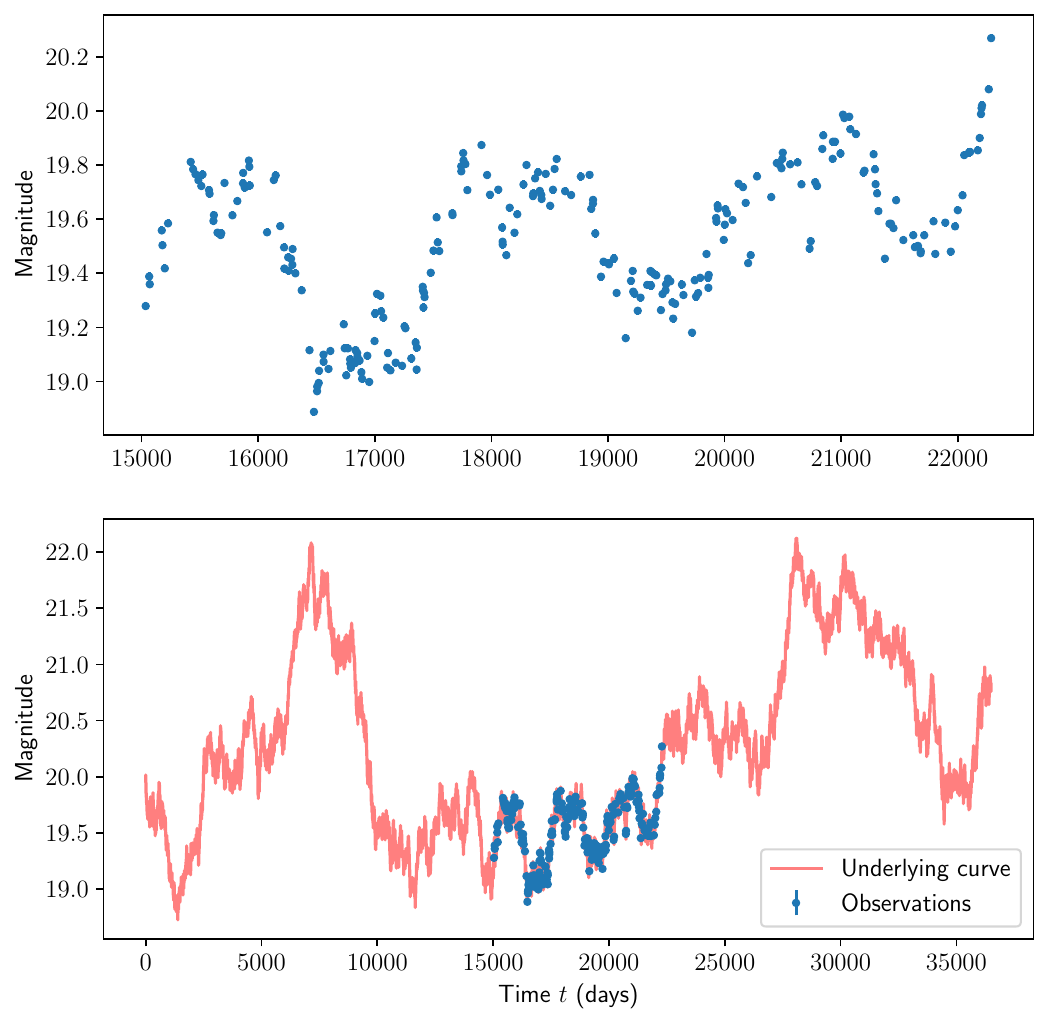}
\end{center}
\caption{Top panel: Simulated light curve measurements
generated from a CAR(1) model with additive gaussian noise.
Naively, the data seem to indicate that $\mu \approx 19.5$ and
$\sigma \approx 0.3$.
Bottom panel: The same simulated measurements (blue)
but zoomed out, showing the true curve (red) from which
the measurements were taken. In reality, the
values of $\sigma$ and $\tau$ are much larger than the
data would naively suggest, and the value of $\mu$ could easily be quite
different from 19.5 (it was in fact 20.2). The true value of
$\tau$ is around 6000 days.
\label{fig:simulation}}
\end{figure}

In this section, we show an example with a simulated light curve
where the true parameter values are known,
to illustrate the difference this decision (keeping $\mu$ free vs.
fixing it at a point estimate) can make.
Figure~\ref{fig:simulation} shows an example of a light curve
$\by$
generated from a CAR(1) process and measured with $N=250$
observations over a 20 year observation period, shown
as blue points. The duration and the number of data points in the
simulated dataset is comparable to the extent of the light curves
in S22.
For this simulated dataset, the true values of the parameters
were
$\{\mu, \sigma, \eta, {\rm jitter}\} = \{20.15, 0.992, 0.0181, 0.0064\}$
with a corresponding timescale of $\log_{10}(\tau) = 3.779$
(about 6000 days or 16.5 years).

\begin{figure}
    \centering

    \begin{minipage}{0.48\textwidth}
        \centering
        \includegraphics[width=\linewidth]{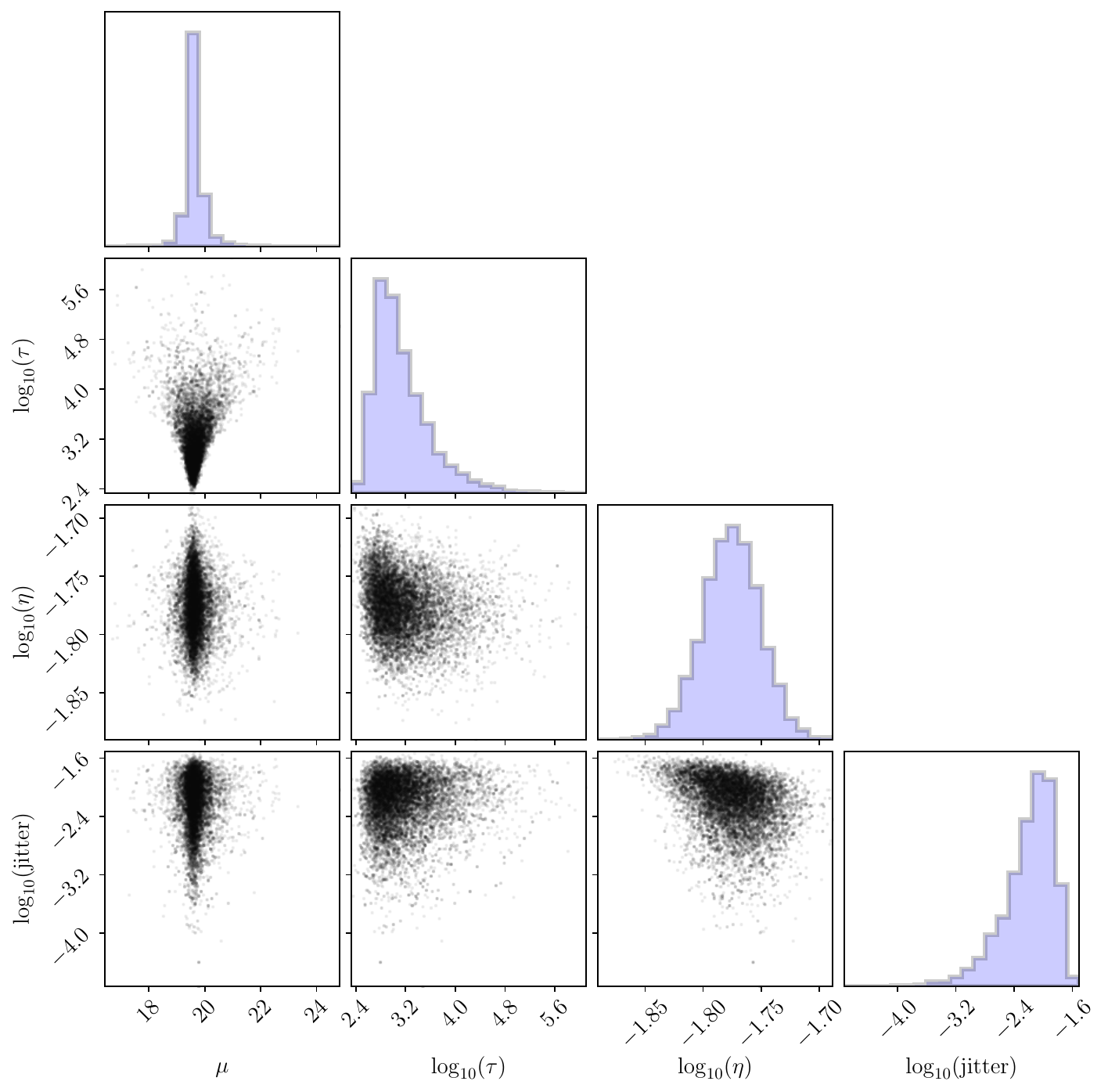}
        \end{minipage}
    \hfill
    \begin{minipage}{0.48\textwidth}
        \centering
        \includegraphics[width=\linewidth]{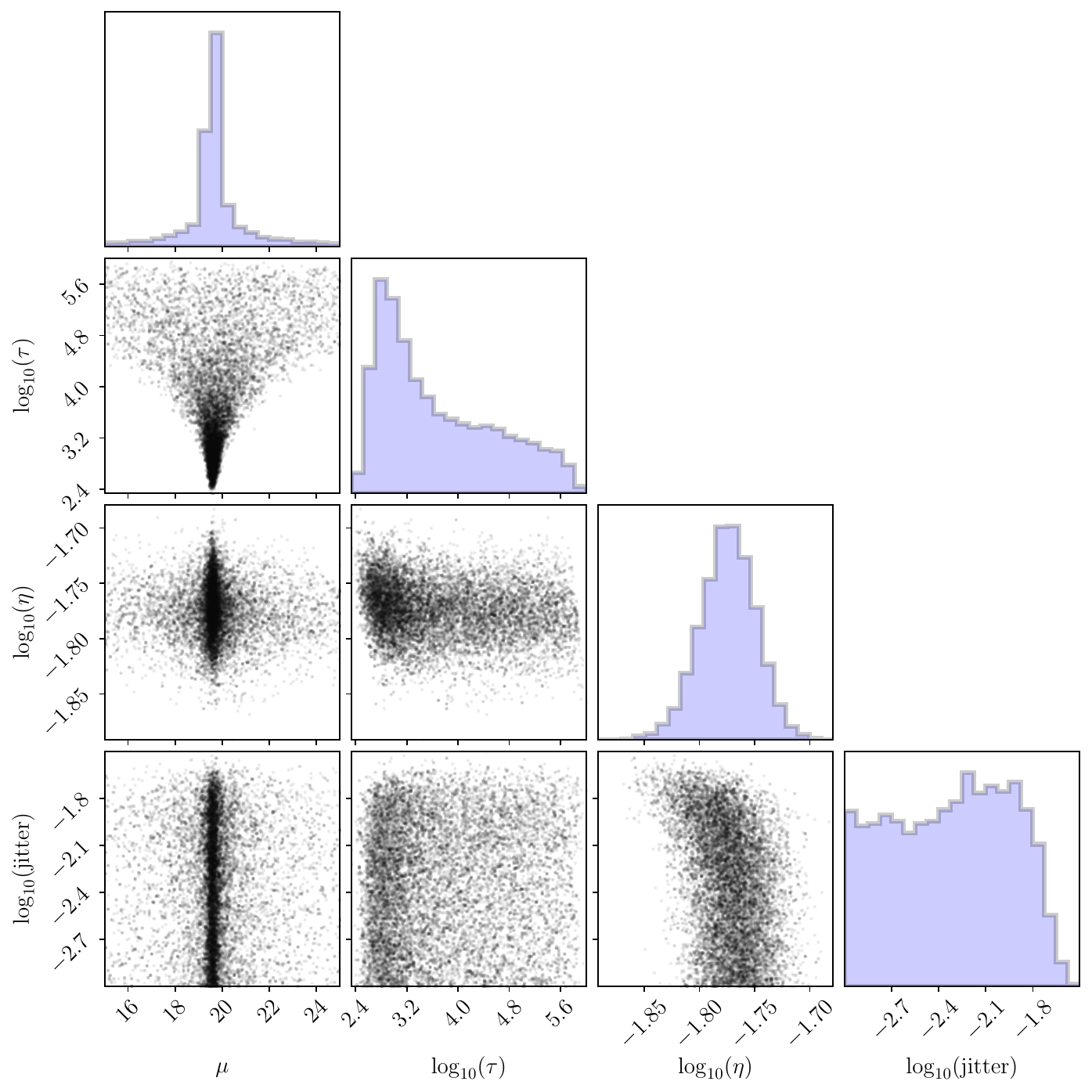}
    \end{minipage}

\caption{Corner plots of the posterior distribution for $\mu$, $\tau$, $\eta$, and the jitter parameter
based on the simulated dataset in Figure~\ref{fig:simulation}.
On the left, the posterior is obtained from the informative priors, and on the right,
from the flat priors.
Due to the funnel shaped joint posterior
for $\mu$ and $\tau$,
The uncertainty about $\tau$ is extremely large, covering a few orders
of magnitude, especially in the case of the flat priors.
Notably, the volatility parameter $\eta$ is well constrained in both cases.
\label{fig:cornerplots}}
\end{figure}

Despite the nontrivial number of measurements in the simulated
dataset, the posterior distribution indicates that it is
compatible with a wide range of possible values of the
parameters, as the marginal posterior distributions are
very broad. Corner plots based on the informative and the
flat priors are shown in Figure~\ref{fig:cornerplots}.
While the computation was
done in the $(\mu, \sigma, \eta, {\rm jitter})$ coordinate system,
we transformed back to $(\mu, \tau, \eta, {\rm jitter})$
for these corner plots, since $\tau$ is of particular interest.
The marginal posterior distributions
for all parameters apart from $\eta$ are very broad. 
The marginal posterior distribution for $\mu$ does peak near
the average of the observations, however, values that depart
from the average of the observations are still somewhat plausible,
especially under the flat priors.
The
joint distributions show that, as $\tau$ increases,
the uncertainty about $\mu$ gets larger (as shown by the
characteristic funnel shape in the joint posterior for
$\mu$ and $\tau$, which is more exaggerated under the flat
priors), and a wide range
of $\sigma$ and $\tau$ values remains plausible.
Indeed, values of $\tau$ much larger than the observation window
remain plausible, and were in fact true when we generated the
data.
Generally speaking, light curves like this one can be explained by moderate
$\tau$ values (of order the duration of the observations, or somewhat less)
and $\mu$ being close to the average of the data. However,
they can also be explained by having a large timescale $\tau$,
a much larger value of $\sigma$ than the data naively suggests,
and a significantly different value of $\mu$ than what the
data naively suggests. This phenomenon is not particular to this
simulated dataset, but occurs over a wide range of parameter space.

\begin{figure}
\includegraphics[width=\textwidth]{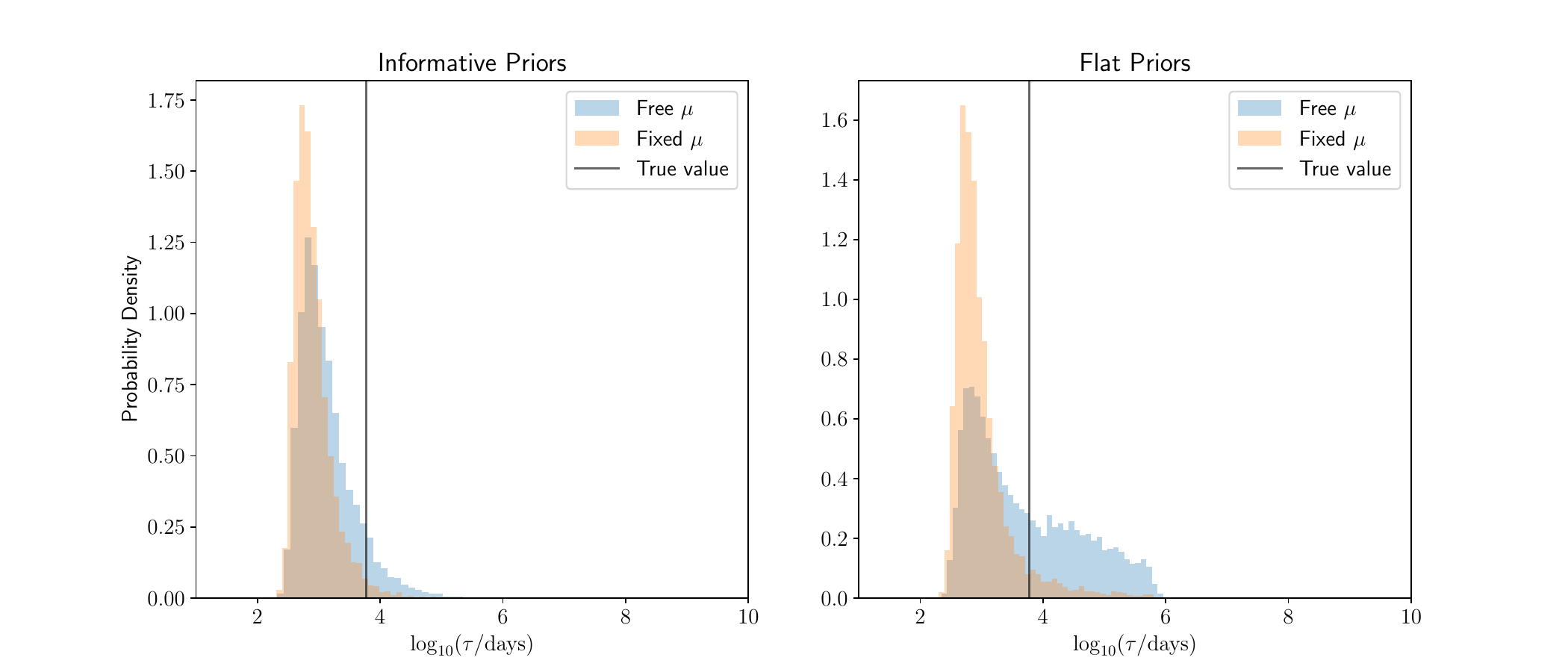}
\caption{The marginal posterior distribution for $\log_{10}(\tau)$ for the simulated
dataset, based on fixing the mean parameter $\mu$, vs. leaving it free.
The true value is indicated by the vertical line. The posterior density at
the true value is, in this case, higher when $\mu$ is allowed to be a free parameter. This is not true for all possible datasets, but is true
in expectation (i.e., averaging over possible datasets), as shown
in Section~\ref{sec:scoring}.
\label{fig:four_posteriors}}
\end{figure}

\begin{figure}
\includegraphics[width=\textwidth]{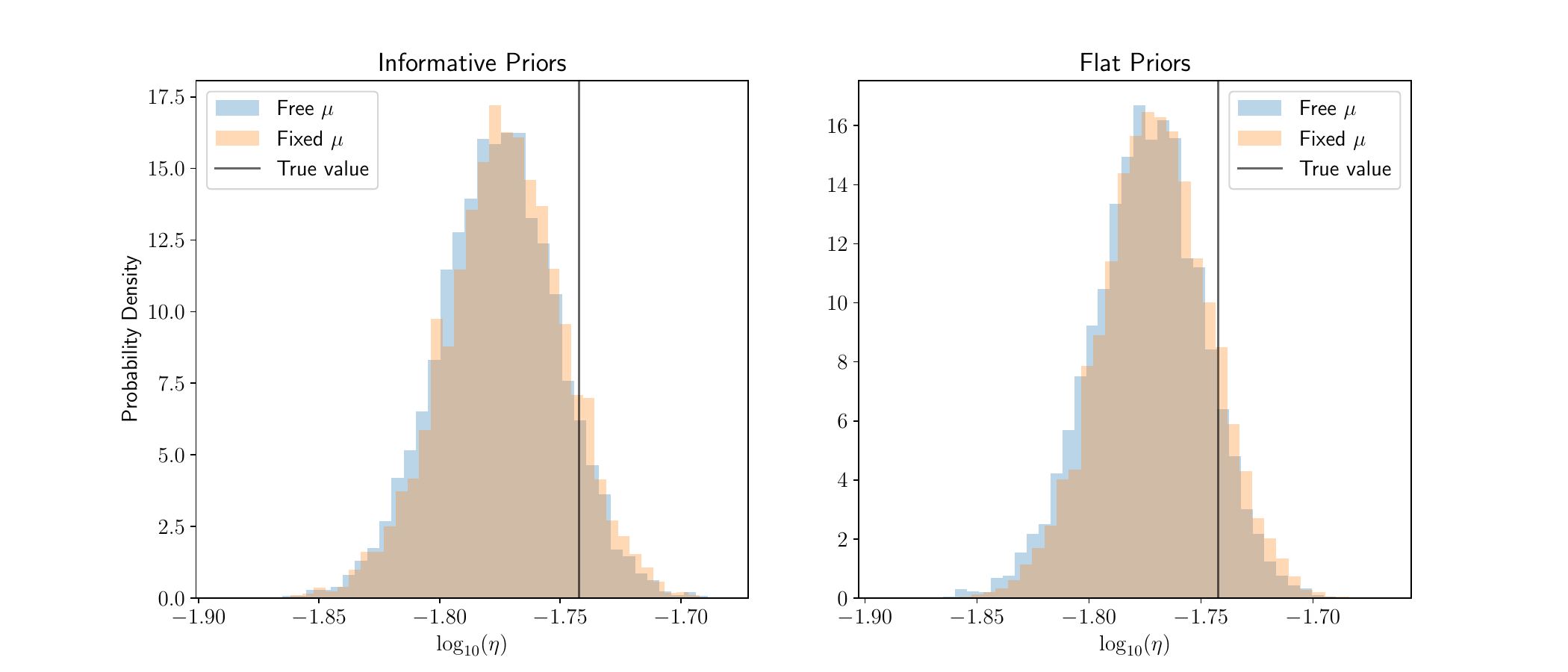}
\caption{The same as Figure~\ref{fig:four_posteriors}, but for the
volatility $\log_{10}(\eta)$ instead of the timescale $\log_{10}(\tau)$.
Under both choices of prior, and under both treatements of the $\mu$
parameter, the posterior distribution is narrow and the density at the
true value is high.
\label{fig:four_posteriors_eta}}
\end{figure}

Fixing $\mu$ to a point estimate (e.g. using Equation~\ref{eqn:fixed_mu}),
rather than keeping it as a free parameter, may appear to resolve the issue
of the broad posterior distribution for $\tau$.
However, this is an artificial resolution, as we do not in fact know that
$\mu = \hat{\mu}$, and we should acknowledge this in our assumptions.
In Figure~\ref{fig:four_posteriors}, the posterior distributions
for $\log_{10}(\tau)$ for the simulated dataset are shown under
four distinct inference settings. In the first panel the informative
priors were used (with the $\mu$ parameter free and then fixed using Equation~\ref{eqn:fixed_mu}),
and in the second panel the flat priors were used
(with the $\mu$ parameter free and then fixed).
As expected, the posterior distributions are narrower when
the informative priors are used (the posterior standard deviations
are 0.46 and 0.31 for the informative priors under free/fixed $\mu$
respectively, and 0.90 and 0.48 for the flat priors under free/fixed
$\mu$ assumptions). However, in 
the case of this simulated dataset, the probability density
at the true value is higher for the free $\mu$ assumption
than for fixed $\mu$ assumption, indicating better performance
(in an ex post sense) when keeping $\mu$ free, at least for this
dataset.
This improved performance from keeping $\mu$ free
would not be universally true across
all possible datasets, but is true on average over possible
datasets, as discussed in Section~\ref{sec:scoring}.
In Figure~\ref{fig:four_posteriors_eta}, we show the equivalent
plot of the posterior distributions for the volatility
$\log_{10}(\eta)$ under the two different priors and the two different
treatments of $\mu$. In all cases, the posterior for
$\log_{10}(\eta)$ is narrow and the probability density at the known
true value is high. The four posterior standard deviations
for $\log_{10}(\eta)$ are all 0.025, to two significant figures.
For completeness, the estimated marginal likelihood for the simulated
dataset under the informative priors (with $\mu$ free) is
$\log(Z) = 377.58$, and for the flat priors it is
{\revision modestly} lower, as expected {\revision from wider priors},
with $\log(Z) = 376.27$. {\revision These are not marginal likelihoods
for competing models and it is not useful to compute the Bayes factor
here. Such a calculation would essentially be an unusual way of
implementing a third prior which is a mixture of the flat and
informative priors.}

\section{Information Theory}\label{sec:information}
In the previous section, we identified two important properties of the
posterior distribution for the CAR(1) parameters
given a particular simulated dataset:
\begin{itemize}
\item [(i)] The marginal posterior distribution for the timescale $\log_{10}(\tau)$,
evaluated at the true value, was higher when $\mu$ was kept free,
compared to when it was fixed at a point estimate.
\item [(ii)] The marginal posterior distribution for the short term
volatility parameter $\log_{10}(\eta)$
was very narrow compared to the marginal posterior distributions for the
other parameters.
\end{itemize}
In this section, we use information theory to test whether these
conclusions generalise to all possible datasets, as defined by the
prior predictive distribution over datasets
\begin{align}
p(\by) &= \int p(\btheta)p(\by \given \btheta) \, d\btheta.
\end{align}
Note that this is the exact equation previously given
for the marginal likelihood (Equation~\ref{eqn:marginal_likelihood}).
This is a quirk of the traditional overloaded notation.
The marginal likelihood is, in fact, the prior predictive distribution
evaluated at the observed data.
This is directly analogous to how the sampling distribution
and the likelihood function
share the same notation, but the latter
is obtained by evaluating the former at the actually observed data.

Information theory can be used within a probabilistic context to quantify
uncertainty and relevance \citep{knuth_questions, KNUTH2005245}. The fundamental quantities
in information theory are entropies and relative entropies (or Kullback-Leibler
(KL) divergences).
For a discrete probability distribution $p(\theta)$,
the Shannon entropy \citep{shannon1948mathematical} is defined as
\begin{align}
H(\theta) = -\sum p(\theta) \log p(\theta),\label{eqn:entropy}
\end{align}
where the sum is over the set of mutually exclusive and exhaustive
possibilities for $\theta$. This expression measures the amount of
uncertainty in the probability distribution. For a discrete uniform
distribution, the entropy is $\log(n)$ where $n$ is the number of
possibilities.
For continuous distributions, which we use throughout this paper,
a continuous version with an integral instead of a sum
(sometimes called differential entropy) can be defined. It is given by
\begin{align}
H(\theta) = -\int p(\theta) \log p(\theta) \, d\theta.
\end{align}
However, in the continuous case the entropy must be understood
with respect to a given coordinate system --- for example,
$H(\theta)$ and $H(\log\theta))$ will usually be different values.
To build intuition, the entropy of a uniform distribution
with volume $V$ is simply $\log(V)$, and the entropy can be regarded
as the generalisation of log-volume to non-uniform distributions.
The fact that the continuous entropy is coordinate dependent arises
from the Jacobian factor that must be included in the volume element
when changing coordinates. However, for our purposes, we will be
concerned with volumes in a fixed coordinate ($\log_{10}(\tau)$
and $\log_{10}(\eta)$) and will not need to worry about any Jacobians.
While the discrete entropy (Equation~\ref{eqn:entropy}) is non-negative,
the continuous version can be negative. The reason is simply that
the number of possibilities in a discrete situation cannot be less than
1, but the volume in a continuous situation can be less than 1.

In a context with an unknown parameter $\theta$ and a particular dataset
$D$, the entropy of the posterior distribution is defined as
\begin{align}
H &= -\int p(\theta \given D)\log p(\theta \given D) \, d\theta.\label{eqn:posterior_entropy1}
\end{align}
However, the conditional entropy, given the symbol $H(\theta \given D)$,
is actually defined by taking the expectation
over datasets,
\begin{align}
H(\theta \given D) &= -\int p(D) \int p(\theta \given D)\log p(\theta \given D) \, d\theta \, dD.\label{eqn:posterior_entropy2}
\end{align}
This quantifies how much uncertainty would remain about $\theta$
if $D$ is taken into account, but without specifying any particular
value for $D$. Whereas Equation~\ref{eqn:posterior_entropy1} is a function
of a question (``what is the value of $\theta$?'') and a statement (``the data
was observed to be $D$''), Equation~\ref{eqn:posterior_entropy2} is a function of
a question (``what is the value of $\theta$?'') and another question
(``what is the value of $D$?'').
Subsequently, we use the term posterior
entropy for Equation~\ref{eqn:posterior_entropy1} and
conditional entropy for Equation~\ref{eqn:posterior_entropy2}.

The mutual information may be defined as
\begin{align}
{\rm MI}(\theta, D) &= H(\theta) - H(\theta \given D).
\end{align}
This is a measure of dependence, which describes
how much learning $D$ tells us about $\theta$.
There are various identities and alternative ways of writing
the mutual information. We will not list them exhaustively,
but note that it is also the Kullback-Leibler divergence
of the joint prior distribution for $\theta$ and $D$
from the product of its marginals:
\begin{align}
{\rm MI}(\theta, D) &= \iint p(\theta, D) \log\left[\frac{p(\theta, D)}{p(\theta)p(D)}\right]
\, d\theta \, dD.
\end{align}
The mutual information
is also the expected value of the prior-to-posterior 
Kullback-Leibler divergence, averaged over possible datasets:
\begin{align}
{\rm MI}(\theta, D) &= \int p(D) \int p(\theta \given D) \log\left[\frac{p(\theta \given D)}{p(\theta)}\right] \, d\theta \, dD.
\end{align}
This second property allows us to estimate the mutual information using
Nested Sampling \citep{skilling}, which can calculate the prior-to-posterior
KL divergence for a particular dataset (Skilling calls this
simply the `information'). However, this approach can only
give the mutual information between the data and the full parameter vector.
If we want to focus on particular parameters such as $\log_{10}(\tau)$
and $\log_{10}(\eta)$, we need a method that also allows for marginalisation
over a subset of the parameters.

Thus, we used a Nested Sampling approach similar to the one introduced
by \citet{brewer2017computing}. Specifically, we used the simpler implementation
provided at \url{https://github.com/eggplantbren/PostEnt2026}, which is designed
to calculate conditional entropies (Equation~\ref{eqn:posterior_entropy2})
rather than arbitrary entropies, and which
can also marginalise over parameters, giving the conditional entropy
for only the parameter(s) of interest.
This implementation also uses Diffusive Nested Sampling for the inner loop,
which is more robust to difficult posterior distributions
than the original approach of \citet{brewer2017computing}.
Switching from generic notation $(\theta, D)$ to the specific
parameters and data for our model $(\btheta, \by)$,
the algorithm simulates parameters $\btheta$ from the
prior and a corresponding dataset $\by$ from the sampling distribution.
Together, these constitute a (parameters, data) pair simulated from the
joint prior distribution $p(\btheta)p(\by \given \btheta)$.
It then estimates the logarithm of the posterior density given $\by$
at the true parameter value, which is known by construction. Finally,
it computes the average over all datasets.
Our simulated datasets for this section are based on $N=250$
observations over a baseline of about 20 years, similar to the
light curves of S22.

\begin{figure}
    \centering

    \begin{minipage}{0.48\textwidth}
        \centering
        \includegraphics[width=\textwidth]{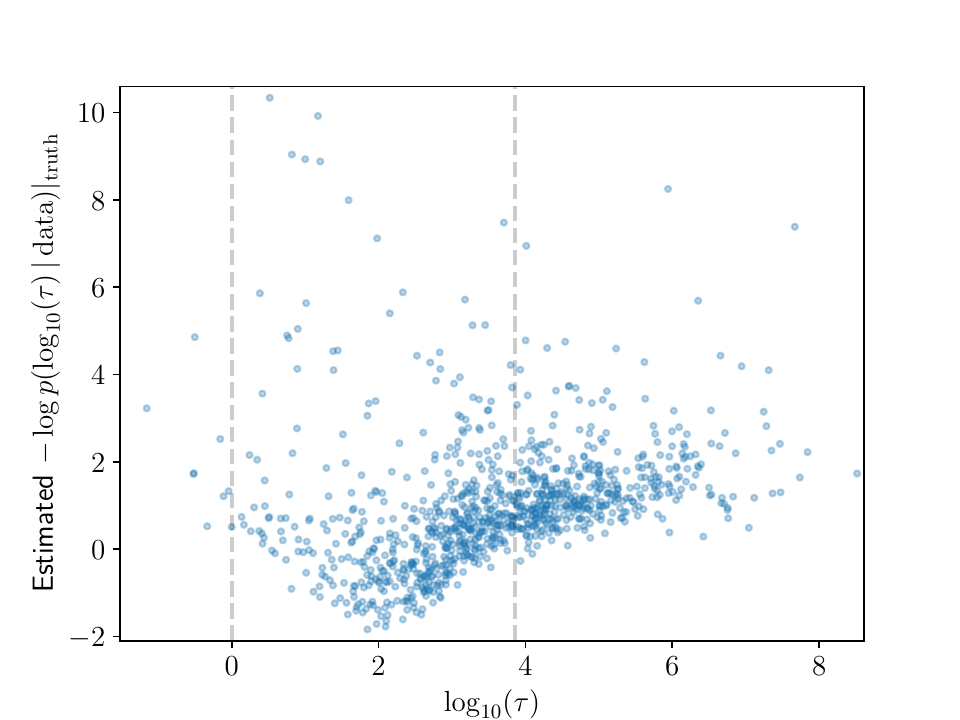}
        \end{minipage}
    \hfill
    \begin{minipage}{0.48\textwidth}
        \centering
        \includegraphics[width=\textwidth]{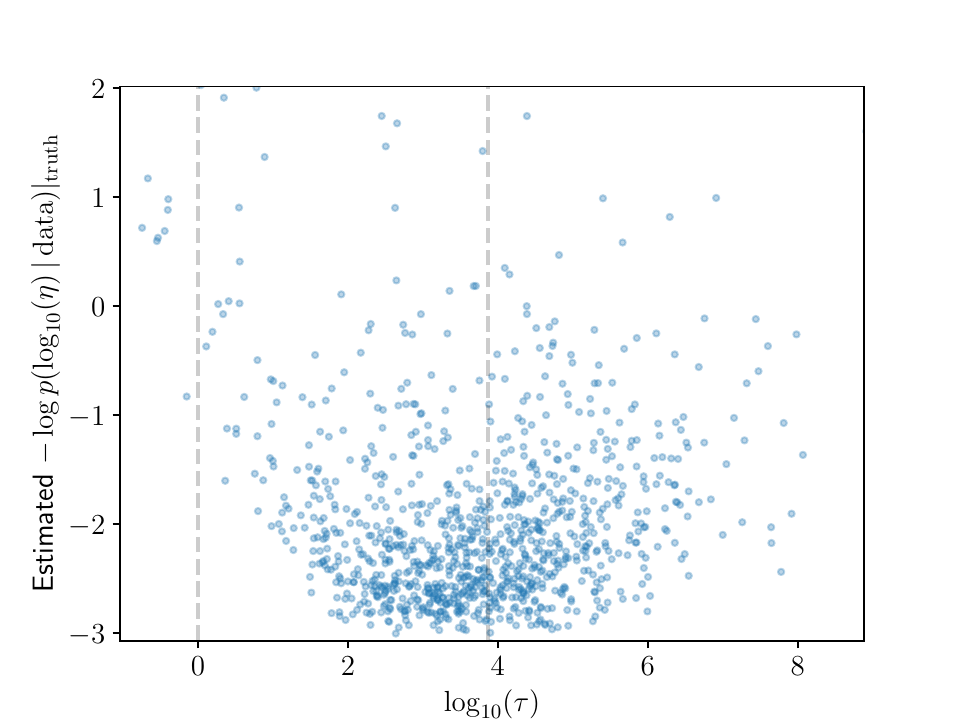}
    \end{minipage}
\caption{For 1000 simulated light curves, we estimated
the entropy of the posterior (on the $y$-axis), and plotted it against
the true timescale ($x$-axis; known because this is based on simulated
data). For $\log_{10}(\tau)$ values above about 2, there is a clear
increasing trend, so it is indeed easier to infer $\tau$
if it is small relative to the duration of the observations.
This is our equivalent of \citet{kozlowski2017limitations}'s
conclusion that $\tau$ should be significantly shorter than
the observation window to be well constrained.
The vertical lines indicate one day (the cadence)
and 20 years (the baseline) respectively.
\label{fig:entropy_vs_tau}}
\end{figure}

Figure~\ref{fig:entropy_vs_tau} shows an estimate of the posterior entropy
for $\log_{10}(\tau)$ as a function of its true value, using the informative
priors. The value plotted on the $y$-axis is minus the estimated log of the
posterior density at the true value, and is part of the calculation process
for $H(\log_{10}(\tau) \given \by)$. Indeed, the conditional entropy is
obtained by averaging all of the $y$-values in the plot.
Between $\log_{10}(\tau)=2$ and $\log_{10}(\tau)=7$, a clear increasing
trend is seen, showing that $\log_{10}(\tau)$ has a narrower posterior
distribution (lower entropy) when $\tau$ is shorter, as expected.
This is a Bayesian, information theoretic version of \citet{kozlowski2017limitations}'s conclusion that
$\tau$ is more constrained when it is short.
Below $\log_{10}(\tau)=2$, the typical posterior entropy becomes higher
again, as short variability timescales mean only an upper limit for $\tau$
is possible, and these posterior distributions do not have a low entropy.  
However, due to $\mu$ being a free parameter in our analysis, the typical
uncertainty is still very large. The conditional entropy of $0.929 \pm 0.070$
nats is roughly equivalent to the posterior being a normal distribution with
a standard deviation of 0.6 dex (making use of the expression
for the entropy of a normal distribution, which is $H = \frac{1}{2}\log(2\pi e \sigma^2)$),
but much of the information is coming
from the prior rather than from the data.
With the flat prior the conditional entropy is $2.063 \pm 0.060$ nats,
corresponding to a normal distribution with a standard deviation of almost 2 dex.

On the other hand, the right hand panel of Figure~\ref{fig:entropy_vs_tau}
shows that $\log_{10}(\eta)$ is typically very well constrained by the
data. The marginal posterior distribution has much higher probability
density at the true value, and the resulting conditional entropy is
much smaller than for $\log_{10}(\tau)$ ($-1.859 \pm 0.054$ vs. $0.929 \pm 0.070$).
For intuition purposes, this conditional entropy is equivalent to a normal
distribution with a standard deviation of about 0.04 dex.
Our observation that $\log_{10}(\eta)$ was much better constrained than
$\log_{10}(\tau)$ indeed generalises from the particular dataset
in Section~\ref{sec:fitting} and is typical across possible datasets
generated from the joint prior distribution over parameters and datasets.

\section{Logarithmic Scoring Rule}\label{sec:scoring}
Scoring rules describe the {\em ex post}
quality of a probability distribution, if the true value of the
unknown quantity later becomes known. Scoring rules are functions
taking the true value and the probability distribution as inputs,
and returning a real number.
The logarithmic scoring rule is the unique local, strictly proper
scoring rule. The term {\em local} means that the score
only depends on the probability
assigned to the true value, not to other values. The term {\em strictly proper} means
that if we believe in a particular probability distribution
$p(\theta)$ then asserting $p(\theta)$ itself uniquely maximises the
expected value of the score \citep{gneiting2007strictly}.
For the informative and the flat priors, the expected logarithmic
score when inferring $\log_{10}(\tau)$ or $\log_{10}(\eta)$
are equivalent to the conditional entropies which we calculated
in the previous section, but with a change of sign.
In Figure~\ref{fig:four_posteriors}, we saw that the posterior density
for $\log_{10}(\tau)$ at the true value was higher if we left $\mu$
as a free parameter, compared to fixing it at a point estimate.
To see whether this property generalises across datasets,
we also computed the expected value of the logarithmic score
in the case where $\mu$ is fixed at a point estimate.
Under the informative priors, when the goal is to infer $\log_{10}(\tau)$,
we find that
the expected logarithmic score is only slightly higher when keeping $\mu$
free ($-0.929 \pm 0.070$) than when fixing $\mu$ at a point estimate
($-0.992 \pm 0.083$); indeed, this difference could be Monte Carlo noise.
However, under the flat priors, the difference becomes much
more dramatic.
Treating $\mu$ as a free parameter gives an expected logarithmic score of $-2.063 \pm 0.060$,
whereas fixing it gives $-2.936 \pm 0.123$, so the posterior density at the
true value is almost a factor of $e$ smaller, on average.

We did not evaluate the fixed-$\mu$ strategy with respect to inference
of the short term volatility parameter
$\log_{10}(\eta)$, as Figure~\ref{fig:four_posteriors_eta} suggests is likely to be insensitive to this choice.
The parameter $\eta$ is the day-to-day variability in magnitudes, and is
well constrained in general. The expected logarithmic scores under the
free-$\mu$ setting (which are just our posterior entropies
from Section~\ref{sec:information}) with the sign
reversed) are significantly higher than for $\log_{10}(\tau)$, as
expected.
We note that, for $\log_{10}(\eta)$, the conditional entropy is significantly
lower for the informative priors than for the flat priors. This is because
the flat priors allow some datasets with very short timescales $\tau$
(Figure~\ref{fig:tau_priors}), for which
the uncertainty for $\log_{10}(\eta)$ becomes quite large.
However, this region of parameter space is not really as plausible as
the flat prior suggests.

A summary table of information theoretic quantities and expected
logarithmic scores is displayed in Table~\ref{tab:information}.
For completeness, we also calculated the mutual informations which
are the differences between the prior and posterior entropies, and quantify
how much information the data provides about the parameter of interest.
\begin{table}
\centering

\begin{minipage}{0.9\textwidth}
\centering
$\log_{10}(\tau)$ \\[0.5em]
\begin{tabular}{|l|l|c|c|c|c|}
\hline
Priors & Treatment of $\mu$ & Prior Entropy & Conditional Entropy & Mutual Information & Expected Logarithmic Score \\
\hline
Informative & Free & 1.957 $\pm$ 0.039 & 0.929 $\pm$ 0.070 & 1.028 $\pm$ 0.080 & -0.929 $\pm$ 0.070  \\
Informative & Fixed & 1.957 $\pm$ 0.039 & & &  -0.992 $\pm$ 0.083 \\
Flat & Free & 2.849 $\pm$ 0.037 & 2.063 $\pm$ 0.060 & 0.786 $\pm$ 0.070 & -2.063 $\pm$ 0.060\\
Flat & Fixed & 2.849 $\pm$ 0.037 & & & -2.936 $\pm$ 0.123 \\
\hline
\end{tabular}
\end{minipage}

\vspace{1em}
\begin{minipage}{0.9\textwidth}\centering
$\log_{10}(\eta)$ \\[0.3em]
\begin{tabular}{|l|l|c|c|c|c|c|}
\hline
Priors & Treatment of $\mu$ & Prior Entropy & Conditional Entropy & Mutual Information & Expected Logarithmic Score \\
\hline
Informative & Free   & 0.891 & -1.859 $\pm$ 0.054 & 2.750 $\pm$ 0.054 &  1.859 $\pm$ 0.054 \\
Flat & Free        & 1.609 & 0.146 $\pm$ 0.111 & 1.463 $\pm$ 0.111 & -0.146 $\pm$ 0.111\\
\hline
\end{tabular}
\end{minipage}

\caption{Table of information theoretic quantities regarding inference
of the timescale
$\log_{10}(\tau)$ and the volatility $\log_{10}(\eta)$ under different
assumptions (informative/flat priors, and free/fixed $\mu$ strategies).
The conditional entropy describes the typical
amount known about $\log_{10}(\tau)$ after conditioning on the data.
All values are based on $N=250$ observations over 20 years.
\label{tab:information}}
\end{table}

\section{Hierarchical Analysis}\label{sec:hierarchical}
Having established that $\tau$ is very difficult to infer when $\mu$
is a free parameter, and that $\mu$ actually {\em should} be a free parameter,
we now turn to the question of what we should do given this reality.
Since the short term volatility $\eta$ is much more well constrained by
light curve data, it is likely
that focusing on $\eta$ instead of $\tau$ could be a fruitful line of investigation.
In this section, we present an initial hierarchical analysis as a proof of
concept of this strategy. We note that in his seminal paper on this topic,
\citet{kelly2009variations} did perform some regressions with $\sigma$
(corresponding to our $\eta$, not our $\sigma$) with respect to luminosity,
black hole mass and Eddington ratio. However, later studies have overwhelmingly
emphasised $\tau$ \citep[e.g.][]{suberlak2021improving, burke2021characteristic}.
This might partially be caused by the increased popularity of the Gaussian process formalism,
where $\sigma$ is usually preferred, over the stochastic differential equation
formalism which in which $\eta$ occurs naturally.
Our analysis differs in that we identify $\eta$ as the only well-constrained
parameter under realistic observing conditions and build a hierarchical
population model specifically for it.

The dataset is from S22 and consists of light curves for 190 quasars,
each measured in three bands ($g$, $r$, and $i$) over a period of 20 years.
We computed the rest-frame wavelengths of the observations in each band
using
\begin{align}
\lambda_g &= \frac{4720 \mathring{\mathrm{A}}}{1+z} \\
\lambda_r &= \frac{6415 \mathring{\mathrm{A}}}{1+z} \\ 
\lambda_i &= \frac{7835 \mathring{\mathrm{A}}}{1+z}.
\end{align}
Previous studies used the $\log_{10}(\tau)$ estimates from S22
as the response variable in a regression model, in order
to detect the expected time dilation signal,
in the form of a term $n\log_{10}(1+z)$, where
$n=1$ implies a pure cosmological time dilation signal with no intrinsic evolution
in quasar properties over time
\citep{lewis2023detection, brewer2025revisiting}. However, here
we will make two changes. Firstly, we will treat $\eta$ as the
response variable in our regression model, not $\tau$.
Secondly, rather than estimating $\eta$ for each light curve and treating
that output as data, we will build a model whose likelihood function
is based on the light curves themselves. While computationally much more intensive,
this approach avoids some philosophical and practical difficulties.
Treating posterior distributions (i.e., inferred $\eta$ or $\tau$ values
and their uncertainties) as if they were `data', or even likelihood functions,
is a subtle issue \citep{2018arXiv180407766H}.
Also, both \citet{lewis2023detection} and \citet{brewer2025revisiting}
had to adopt unconventional asymmetric probability distributions
to deal with the asymmetric uncertainties on $\log_{10}(\tau)$
reported by S22. Using the raw light curves means that
our results will reflect the actual shape of the likelihood function,
rather than an approximate shape reconstructed from
published credible intervals. However, since the posterior distributions
for $\log_{10}(\eta)$ from individual light curves are usually
narrow normal distributions (unlike posterior distributions for
$\log_{10}(\tau)$), this choice is unlikely to have a major numerical
impact on the results.

\subsection{Data Cleaning}
Some of the light curves in the S22 dataset, largely in the $r$ and $i$ band
light curves, contain discrepant
magnitude measurements which are clearly unreliable, with
a reported unphysical magnitude of $-44$. Such points 
must be discarded, but there are other points which
are less extreme which may not be accurate as they suddenly
deviate from the overall trend of the light curve.
These points have two main effects on the posterior distributions
for the CAR(1) parameters. The model has two main ways of
explaining how such a discrepant point could have occurred.
The first is by increasing the jitter term and explaining the
discrepant point as a white noise fluctuation. The second is
to reduce the value of $\tau$, which has a similar effect,
making the CAR(1) process more similar to a white noise
distribution.
Before running the inferences, the previous analyses
ran the light curves through a Hampel filter \citep{hampel}
to remove outliers
(Stone, priv. comm).
We adopted a simpler approach of discarding any measurement
that was more than 5 times the interquartile range away
from the median value of the light curve.
Using this approach, a total of 18 outliers were removed,
across 13 out of the 570 light curves.
A more principled treatment of these discrepant measurements
could be developed involving a heavy-tailed noise model, however,
such an approach is beyond the scope of this paper.

\subsection{Model Specification}
Our hierarchical model aims to simultaneously infer the four CAR(1) quantities
for each of the $190 \times 3$ light curves:
\begin{align}
\left(\mu_{ij}, \log_{10}(\sigma_{ij}), \log_{10}(\eta_{ij}),
\log_{10}(\textnormal{jitter}_{ij})\right),
\end{align}
where $i \in \{1, ..., 190\}$ indexes the quasars and
and $j \in \{g, r, i\}$
indexes the waveband of the observations.
We use hierarchical priors, so that the inference is allowed to
borrow strength across light curves --- for instance, if the
data suggests $\log_{10}(\sigma)$ clusters around a typical value
for many light curves,
that clustering will provide information that is relevant
to inference of $\log_{10}(\sigma)$ for the other light curves.
This requires the introduction of hyperparameters, describing concepts
such as `the typical value of $\mu$ across all light curves',
the `diversity of $\mu$ across all light curves', and so on.

The entire model specification including priors for the hyperparameters,
the parameters, and the data, is given in Table~\ref{tab:hierarchical_priors}.
There are 11 hyperparameters and 2280 ($=190\times3\times4$) parameters,
for a total of 2291 unknowns.
We use $m$ to denote hyperparameters that are means and $\omega$ for hyperparameters
that are standard deviations, with subscripts to indicate which quantity
the mean and standard deviation applies to (e.g., $m_{\mu}$ and $\omega_{\mu}$
describe the mean and standard deviation of magnitude parameters $\mu$ across
all light curves).
The prior for all individual light curve parameters is based on the appropriate
hyperparameters, and we use normal distributions for this purpose.
However, $\eta$ is treated differently from the other parameters, and we
apply a regression assumption for the expected value of $\log_{10}(\eta)$,
based on its expected dependence on
explanatory variables, namely the bolometric luminosity $L_{\rm bol}$ of the
quasar (in ergs per second),
the rest wavelength $\lambda$ of the light curve (in Angstroms),
and the redshift $z$ of the quasar. 
The regression model is based on centered explanatory variables,
given by
\begin{align}
\lambda'_{ij} &= \log_{10}(\lambda_{ij}) - \overline{\log_{10}(\lambda)}\\
L'_{{\rm bol}, i} &= \log_{10}(L_{{\rm bol}, i}) - \overline{\log_{10}(L_{\rm bol})}\\
z'_{i} &= \log_{10}(1+z_i) - \overline{\log_{10}(1+z)},
\end{align}
where the overbar denotes the arithmetic mean.
This choice is traditional in Bayesian regression modelling
and tends to make the posterior distribution for the coefficients
more independent (improving computational efficiency), and also changes
the interpretation of $\beta_0$, which becomes the typical value for
$\log_{10}(\eta)$ at a typical bolometric luminosity, rest wavelength, and
redshift. The numerical values of the arithmetic means for the actual data
are
\begin{align}
\overline{\log_{10}(\lambda)}&= 3.378\\
\overline{\log_{10}(L_{\rm bol})} &= 46.16\\
\overline{\log_{10}(1+z)} &= 0.4134.
\end{align}
The expected value of the log of the volatility, $\log_{10}(\eta)$,
for the regression model is then given by
\begin{align}
\beta_0
+\beta_1\lambda'_{ij} + \beta_2L'_{{\rm bol}, i} + \beta_3z'_i.
\end{align}
In this proof of concept analysis, we do not perform variable selection or explore additional explanatory variables such as black hole mass or Eddington ratio, nor do we include nonlinear terms in the regression surface. While scientifically interesting, these extensions would require a substantially more complex hierarchical model --- especially if accounting for uncertainty in the explanatory variables --- and would dilute the focus of this study.

\subsection{Priors}
The priors for all hyperparameters were chosen to be uniform distributions for
simplicity. The limits of the uniform distributions for the $m$ hyperparameters,
describing typical values of $(\mu, \sigma, \eta, {\rm jitter})$ across the sample,
correspond to the limits
in the flat priors described in Section~\ref{sec:fitting}. For example,
we previously set the prior for $\mu$ for an single light curve to be Uniform$(15,25)$,
and now we use that same prior for $m_{\mu}$, the typical value of $\mu$ across
all light curves. The same logic has been applied to the $m$ hyperparameters
for the typical values of $\log_{10}(\sigma)$, $\log_{10}(\eta)$,
and $\log_{10}({\rm jitter})$.

For the regression coefficients $\beta_0$, $\beta_1$, and $\beta_2$, we set
the prior limits by considering their effect on the response variable
$\log_{10}(\eta)$. The range of all explanatory variables ($\log_{10}(L_{\rm bol})$,
$\log_{10}(\lambda)$, and $\log_{10}(1+z)$) is roughly unity,
so our prior limits of $\pm 10$ imply that these could cause variations of
about $\pm 10$
in the response variable $\log_{10}(\eta)$.
Thus, these bounds are generously wide
and should not be increased as this would simply include more extreme implausible
values in the prior distribution.
The lower limit for standard deviation hyperparameters $\omega$
was set to zero and the upper limits were set to 5. Realistically, lower
values are more plausible and values approaching 5 are extremely implausible,
as they would imply that the individual light curve parameters are scattered
across a massive range. However, we did not attempt to incorporate this kind
of prior judgment into the analysis explicitly.
More elaborate hierarchical structures --- for example, allowing 
$\mu$ or $\sigma$
to depend on quasar properties --- are certainly possible,
but we regard them as beyond the scope of this work, whose aim is specifically to
investigate how the volatility $\log_{10}(\eta)$ may depend on these quantities.

Finally, the likelihood function is given by the product of the Gaussian Process
likelihood (Equation~\ref{eqn:gp}) over the 570 light curves.

\begin{table}
\centering
\begin{tabular}{|l|l|l|}
\hline
Quantity & Meaning & Prior \\
\hline
Hyperparameters & & \\
\hline
$m_\mu$      & Typical value of $\mu$ across light curves & Uniform$(15, 25)$\\
$\omega_\mu$      & Diversity of $\mu$ across light curves & Uniform$(0, 5)$\\
$m_{\log_{10}(\sigma)}$      & Typical value of $\log_{10}(\sigma)$ across light curves & Uniform$(-3, 1)$\\
$\omega_{\log_{10}(\sigma)}$      & Diversity of $\log_{10}(\sigma)$ across light curves & Uniform$(0, 5)$\\
$\beta_0$    & Typical value of $\log_{10}(\eta)$ across light curves & Uniform$(-5, 5)$\\
$\beta_1$    & Coefficient of $\log_{10}(\lambda)$ in predicting $\log_{10}(\eta)$ & Uniform$(-10, 10)$\\
$\beta_2$    & Coefficient of $\log_{10}(L_{\rm bol})$ in predicting $\log_{10}(\eta)$ & Uniform$(-10, 10)$\\
$\beta_3$          & Coefficient of $\log_{10}(1+z)$ in predicting $\log_{10}(\eta)$ & Uniform$(-3, 3)$\\
$\omega_{\log_{10}(\eta)}$ & Diversity of $\log_{10}(\eta)$ around regression model & Uniform$(0, 5)$\\
$m_{\log_{10}({\rm jitter})}$ & Typical value of $\log_{10}({\rm jitter})$ across light curves & Uniform$(-3, -1)$\\
$\omega_{\log_{10}({\rm jitter})}$ & Diversity of $\log_{10}({\rm jitter})$ across light curves & Uniform$(0, 5)$\\
\hline
Parameters & & \\
\hline
$\mu_{ij}$ & Mean magnitude of quasar $i$ in band $j$ & Normal$(m_\mu, \omega_\mu^2)$\\
$\sigma_{ij}$ & Variability amplitude of quasar $i$ in band $j$ & $\log_{10}(\sigma_{ij}) \sim \textnormal{Normal}\left(m_{\log_{10}(\sigma)}, \omega_{\log_{10}(\sigma)}^2\right)$ \\
$\eta_{ij}$ & Short term volatility of quasar $i$ in band $j$ & $\log_{10}(\eta_{ij}) \sim \textnormal{Normal}\left(\beta_0
+\beta_1\lambda'_{ij} + \beta_2L'_{{\rm bol}, i} + nz'_i, \omega_{\log_{10}(\eta)}^2\right)$\\
${\rm jitter}_{ij}$ & Jitter of quasar $i$ in band $j$ & $\log_{10}({\rm jitter}_{ij}) \sim \textnormal{Normal}\left(m_{\log_{10}({\rm jitter})}, \omega_{\log_{10}({\rm jitter})}^2\right)$\\
\hline
Data & & \\
\hline
$\by_{ij}$ & Light curve for quasar $i$ in band $j$ & GaussianProcess$(\mu_{ij}, \sigma_{ij}, \eta_{ij}, {\rm jitter}_{ij})$ (Equation~\ref{eqn:gp}) \\
\hline
Prior Information & & \\
\hline
$\boldsymbol{t}_{ij}$ & Timestamps for quasar $i$ in band $j$ & Values known \\
$\boldsymbol{s}_{ij}$ & Error bars for quasar $i$ in band $j$ & Values known \\
$L_{{\rm bol}, i}$ & Bolometric luminosity for quasar $i$ & Values known\\
$\lambda_{ij}$ & Rest wavelength of quasar $i$ in band $j$ & Values known\\
$z_i$ & Redshift of quasar $i$ & Values known \\
\hline
\end{tabular}
\caption{The complete definition of our hierarchical regression model
which aims to predict $\log_{10}(\eta)$ as a function of
rest wavelength, bolometric luminosity, and redshift.
For all hyperparameters, uniform priors were used for simplicity.
The limits on the uniform priors are chosen to be represent physically plausible
values.
\label{tab:hierarchical_priors}}
\end{table}

\subsection{Results}
\begin{figure}
    \centering
        \includegraphics[width=\textwidth]{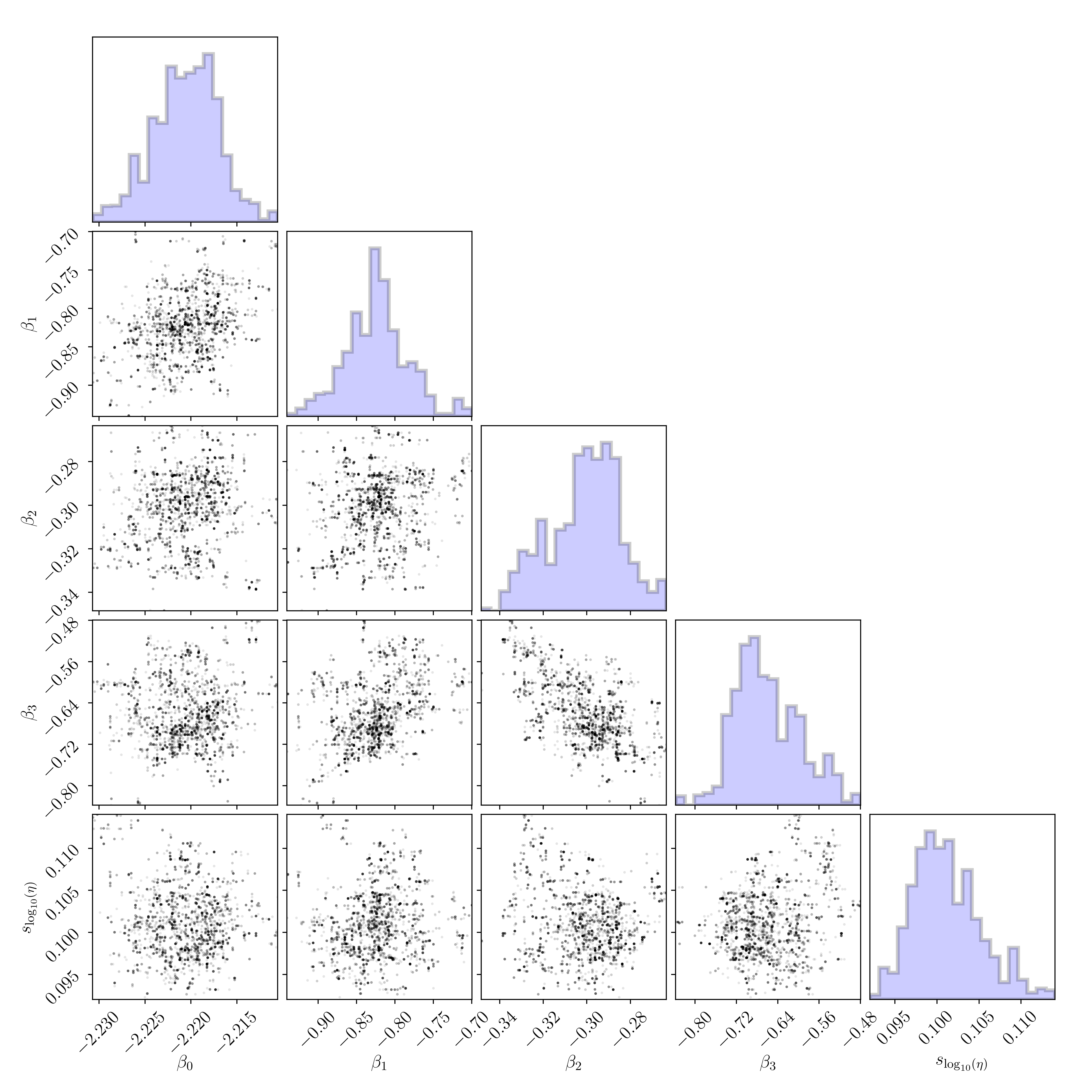}
\caption{A corner plot of a subset of the hyperparameters of the
hierarchical model, particularly, those hyperparameters describing the
relationship between the explanatory variables (rest wavelength and
bolometric luminosity) and the quasar volatility. As a reminder,
$\beta_0$ is the baseline value of $\log_{10}(\eta)$, $\beta_1$
and $\beta_2$ are the coefficients of $\log_{10}(\lambda)$ and
$\log_{10}(L_{\rm bol})$ respectively, and $\beta_3$ is the coefficient
of $\log_{10}(1+z)$. Finally, $s_{\log_{10}(\eta)}$ describes the
dispersion of volatility values around the regression surface.
\label{fig:hierarchical_corner}}
\end{figure}

Posterior summaries for the 11 hyperparameters are presented in
Table~\ref{tab:hierarchical_summaries}. All of the marginal posterior
distributions for the hyperparameters were approximately normal, so we
simply summarised them using the mean $\pm$ the standard deviation.
A corner plot for some of the hyperparameters is also shown in
Figure~\ref{fig:hierarchical_corner}.

\begin{table}
\centering
\begin{tabular}{|l|l|r|}
\hline
Hyperparameter & Meaning & Value \\
\hline
$m_\mu$      & Typical value of $\mu$ across light curves & $19.898 \pm 0.033$\\
$\omega_\mu$      & Diversity of $\mu$ across light curves &  $0.364 \pm 0.062$\\
$m_{\log_{10}(\sigma)}$      & Typical value of $\log_{10}(\sigma)$ across light curves & $-0.341 \pm 0.042 $\\
$\omega_{\log_{10}(\sigma)}$      & Diversity of $\log_{10}(\sigma)$ across light curves & $0.242 \pm 0.018$\\
$\beta_0$    & Typical value of $\log_{10}(\eta)$ across light curves & $-2.2206 \pm 0.0035$\\
$\beta_1$    & Coefficient of $\log_{10}(\lambda)$ in predicting $\log_{10}(\eta)$ & $-0.824 \pm 0.041$\\
$\beta_2$    & Coefficient of $\log_{10}(L_{\rm bol})$ in predicting $\log_{10}(\eta)$ & $-0.301 \pm 0.016$\\
$\beta_3$          & Coefficient of $\log_{10}(1+z)$ in predicting $\log_{10}(\eta)$ & $-0.660 \pm 0.064$\\
$\omega_{\log_{10}(\eta)}$ & Diversity of $\log_{10}(\eta)$ around regression model & $0.1012 \pm 0.0041$\\
$m_{\log_{10}({\rm jitter})}$ & Typical value of $\log_{10}({\rm jitter})$ across light curves & $-2.630 \pm 0.057$\\
$\omega_{\log_{10}({\rm jitter})}$ & Diversity of $\log_{10}({\rm jitter})$ across light curves & $0.707 \pm 0.052$\\
\hline
\end{tabular}
\caption{Posterior summaries for the 11 hyperparameters of the hierarchical model.
The meaning column is duplicated
from Table~\ref{tab:hierarchical_priors} for convenience.\label{tab:hierarchical_summaries}}
\end{table}

The typical value of the mean magnitude $\mu$ is estimated to be
$m_\mu = 19.898 \pm 0.033$, and the diversity hyperparameter for the mean magnitudes
is $\omega_\mu = 0.346 \pm 0.062$. These probably reflect selection effects in how
quasars come to be included in the sample, as opposed to being a literal
description of the properties of all quasars.

The typical value of $\log_{10}(\sigma)$ across light curves is estimated to
be $m_{\log_{10}(\sigma)} = -0.341 \pm 0.042$, and the diversity is
$\omega_{\log_{10}(\sigma)} = 0.242 \pm 0.018$. Interestingly, the aggregated information from
all light curves in the sample has constrained both of these hyperparameters to low values. Presumably, if quasars really did show massive variability
that only shows up over very long timescales,
we would have seen large excursions in {\em some} of the light curves in the sample --- but we do not.
This also provides a path towards inferring $\tau$, if it remains of particular
scientific interest. However, instead of trying to infer it for individual
light curves (which provide very little information about $\tau$),
many light curves can be used simultaneously, and the inference about
$\tau$ for light curve $i$ will also be informed by the data for light
curve $j \neq i$, in the usual hierarchical fashion.
The idea of using hierarchical models to pool information across objects,
where individual objects' properties are not well constrained,
has been explored previously in many astronomical fields, such
as extrasolar planets \citep{hogg2010inferring} and
reverberation mapping \citep{brewer2014hierarchical}.
As a simple demonstration, we computed the posterior distribution
over the arithmetic mean and sample standard deviation of all
570 $\log_{10}(\tau)$ values, which can be derived from the model
parameters. The arithmetic mean is $4.051 \pm 0.074$ (corresponding
to about 31 years) and the standard deviation
is $0.620 \pm 0.032$.

The hyperparameter $\beta_0$, which describes the typical volatility
$\log_{10}(\eta)$ in magnitudes per day$^{1/2}$, is estimated to be
$\log_{10}(\eta) = -2.2206 \pm 0.0035$. However,
since we centered the explanatory variables, $\beta_0$ corresponds to the
typical value of $\log_{10}(\eta)$ at a typical redshift (1.59).
To compute the typical value of $\log_{10}(\eta)$ in the rest frame,
we use
\begin{align}
\beta_0' &= \beta_0 - \beta_3\ \overline{\log_{10}(1+z)} \\
         &= \beta_0 - 0.4134\beta_3,
\end{align}
and the result is $\beta_0' = -1.948 \pm 0.026$. This implies that,
in the rest frame, quasars typically
vary in magnitude by about $0.011$ magnitudes per day$^{1/2}$.

The coefficients $\beta_1$ and $\beta_2$ are both negative, indicating that
volatility decreases as a function of both rest wavelength (i.e., quasars
are more volatile at the blue end of the spectrum) and bolometric luminosity
(i.e., more luminous quasars are less volatile).
The luminosity trend may reflect the fact that more luminous quasars host larger
accretion disks, so that short‑timescale fluctuations are averaged over a larger
emitting area. 

Due to the way $\eta$ appears in the stochastic differential equation
form of the definition of the CAR(1) process, the expected dependence
on redshift, assuming time dilation only, is $\beta_3=-0.5$. Any deviation
from this value implies evolution in the intrinsic properties of quasars.
Our inferred value is $\beta_3 = -0.660 \pm 0.064$, which is slightly lower than $-0.5$.
This suggests that intrinsic volatility decreases with redshift,
implying that quasars have tended to become more volatile (higher $\eta$)
over cosmic time. Due to Equation~\ref{eqn:eta}, there is a factor of $-2$
to convert from our parameter $\beta_3$ to the $\tau$ scaling parameter
(which \citet{lewis2023detection} and \citet{brewer2025revisiting} called
$n$). This factor makes a quasar with intrinsic timescale $\tau$
appear to us as a timescale $\tau (1+z)^n$. When expressed in terms of
$n$, the result is $n=1.32 \pm 0.13$, somewhat higher than the value in the aforementioned papers, which found $n=1.28^{+0.28}_{-0.29}$ and $n=1.14 \pm 0.34$
respectively.
Our result thus indicates intrinsic evolution that was not seen in the S22 $\tau$
estimates, but can be picked up from $\eta$. Our uncertainty is also significantly
smaller, since $\eta$ is much more measurable for individual light curves than
$\tau$.

The $\omega_{\rm \log_{10}(\eta)}$ hyperparameter is an intrinsic scatter
term, describing how much the
log-volatility values tend to depart from the values predicted by the
regression surface. Its value is estimated to be $0.1006 \pm 0.0031$.
This may be compared with
the actual scatter of $\log_{10}(\eta)$ across all light curves, which
is a derived parameter from the model (it is numerically the standard
deviation of all of the $\log_{10}(\eta)$ parameters and can be computed
once for each posterior sample). The squared ratio of these quantities
indicates that the proportion of the variance of $\log_{10}(\eta)$ that is explained
by $\lambda$ and $L_{\rm bol}$ is about 67\%.

The inferred jitter hyperparameters are $m_{\log_{10}({\rm jitter})} = -2.630 \pm 0.057$
and $\omega_{\log_{10}({\rm jitter})} = 0.707 \pm 0.052$, indicating that jitter is a small
but consistent contributor to the noise in light curve measurements.

Finally, for completeness,
the estimated marginal likelihood for this model is $\log(Z) = 238181.7$,
and the prior-to-posterior Kullback-Leibler divergence is about $1650$ nats.
In future studies, alternative models can be compared with this one
without having to redo all of the computation for this model.

\section{Conclusion}
\label{sec:conclusion}
In this paper, we investigated the information theoretic properties of the problem
of inferring stochastic time series parameters from light curve data,
with a focus on the simplest damped random walk commonly used for
quasar variability studies.
The primary issue is the difficulty of inferring the characteristic
timescale $\tau$ without making an artificial assumption that the mean
parameter $\mu$ is known. The variations that we see in typical light
curves can often be explained with extremely large timescales, and potentially
with the mean parameter $\mu$ being very different from the magnitudes
actually observed in the data.
With $\mu$ as a free parameter,
the remaining uncertainty about $\log_{10}(\tau)$
in the posterior distribution, based on a light curve with $N=250$ measurements
over about 20 years, is about $\pm~0.6$ dex with informative priors
and about $\pm~2$ dex with flat priors, which is substantial.
Use of structure functions or other descriptive statistics (such
as a periodogram used as a general power spectral density estimator) does not
resolve the issue about $\mu$, because they implicitly assume that
$\mu$ is known \citep[or, in the case of periodograms, can be marginalised out
assuming a sinusoidal signal;][]{bretthorst1988excerpts}, and therefore inherit the same limitations.
However,
we identified the parameter that is actually well measured by light curve
data --- the short term volatility $\eta$. This parameter is informed by
changes in magnitude observed over short periods (e.g., over days), and
it is straightforward to accumulate such measurements over realistic
observing periods.

Based on this finding, we created a hierarchical regression model to infer
the parameters of the \citet{stone} sample of quasar light curves. The regression model
predicts volatility as a function of rest wavelength, bolometric luminosity,
and redshift. The volatility shows a decreasing trend with respect to
all of these variables. Thus, quasars are more volatile at short wavelengths,
lower bolometric luminosities, and at lower redshift. The redshift dependence
is steeper than cosmological time dilation would predict on its own, indicating
that quasars have evolved to become less volatile over cosmic time.
While our hierarchical model also allows for improved inference of $\tau$
by pooling information across light curves, these inferences should be interpreted
with caution as it is likely that the CAR(1) model is misspecified, i.e., it
makes unrealistic predictions over very long timescales. Thus, the `measured'
$\tau$ values from the hierarchical model, while much more constrained than those from single light curves, may not be physically meaningful.

{\revision The use of some related models, such as CARMA models
\citep{kelly2014flexible} is likely to be affected
by the same issues identified for the CAR(1) model in this paper.
The key issue is whether the autocorrelation timescales could
potentially be comparable to or longer than the observations,
making a point estimate of the mean parameter, $\hat{\mu}$,
potentially inaccurate.
If such long timescales are plausible, the mean parameter should
not be fixed at a point estimate in order for the full uncertainties
to be honestly represented in the posterior distribution.
In this paper, we used a simple constant mean function $\mu$
throughout. If more sophisticated mean functions are used
\citep[e.g.][]{kroupa2026time}, similar issues may occur
if tight priors are used for any hyperparameters controlling
the typical level of the mean function.

Conversely, stochastic oscillation models such as those used
in asteroseismology \citep[e.g.][]{brewer2009gaussian} and for stellar
rotation \citep[e.g.][]{angus2018inferring} are unlikely to be
significantly affected as such oscillations are usually far
shorter than the observation window and the autocorrelation function
will have decayed to approximately zero over such a time period.
}

While there are significant difficulties involved in inference for the CAR(1)
{\revision and related models},
some fields which use these models will not be affected by these
difficulties.
For instance, reverberation mapping \citep[e.g.][]{2011ApJ...735...80Z, pancoast2011geometric}
and gravitational lens time delays \citep{hojjati2014next, kj4s-nnyx},
which use stochastic variability
models primarily to interpolate and extrapolate light curves over
short time periods,
should not be significantly
affected by the arguments presented in this paper.
These methods are mostly sensitive to the
uncertainty in the details of $y(t)$, rather than the stochastic
model parameters (which are hyperparameters in this context).
{\revision The uncertainty in the extrapolated signal $y(t)$
over a short time period
will not be changed much depending on the assumptions made about
$\mu$.}

We conclude that {\revision accounting for the methodological issues
raised in this paper should lead to more robust quasar variability
studies in the future.}

\section*{Acknowledgments}
It is a pleasure to thank Tommaso Treu (UCLA) for a helpful conversation about
this topic.
Will Farr (Stony Brook, Flatiron) and
David Hogg (NYU, Flatiron) also provided input about an approximation scheme
for hierarchical models that we used in some preliminary analyses, and in Ryan Yu's honours dissertation, but
did not make it into the final paper.
{\revision We also thank Namu Kroupa (KICC, Cambridge)
and the anonymous referee
for their comments on the submitted version of the paper.}

\section*{Software \& Data Availability}
The software implementing the analyses presented in this paper is
available online under the MIT licence at
\url{https://github.com/eggplantbren/NewCAR}.
All model fitting is performed using Diffusive Nested Sampling
\citep{dns, dnest4}, a variant of Nested Sampling \citep{skilling} that
uses Markov Chain Monte Carlo (MCMC) exclusively.
To compute information theoretic quantities
such as posterior entropies and mutual informations presented
in Section~\ref{sec:information}, we used a modified version
of the algorithm by \citet{brewer2017computing}, which is implemented
at \url{https://github.com/eggplantbren/PostEnt2026}.

\bibliographystyle{apalike}

\bibliography{oja_template}

%
%

\end{document}